\documentclass[conference]{IEEEtran}

\usepackage{amsmath}
\usepackage[tight]{subfigure}
\usepackage[footnotesize]{caption} 
\usepackage{cite}
\usepackage{eurosym}
\usepackage{amsfonts}
\usepackage{graphics}
\usepackage{epsfig}
\usepackage{psfrag}
\usepackage{pstricks}
\usepackage{array}
\usepackage{shortvrb}
\usepackage{epsf}
\usepackage{graphicx}
\usepackage{rotating}
\usepackage{float}
\usepackage{color}
\usepackage{cite}
\usepackage{soul}
\usepackage{multirow}
\usepackage{acronym}

\acrodef{ess}[\textsc{ess}]{Energy Storage System}
\acrodef{dae}[\textsc{dae}]{Differential Algebraic Equation}
\acrodef{vsc}[\textsc{vsc}]{Voltage Sourced Converter}
\acrodef{coi}[\textsc{coi}]{Centre of Inertia}
\acrodef{fdf}[\textsc{fdf}]{Frequency Divider Formula}
\acrodef{wecs}[\textsc{wecs}]{Wind Energy Conversion System}
\acrodef{spvg}[\textsc{spvg}]{Solar Photo-Voltaic Generation}
\acrodef{tcl}[\textsc{tcl}]{Thermostatically Controlled Load}
\acrodef{hvdc}[\textsc{hvdc}]{High-Voltage Direct Current}
\acrodef{pll}[\textsc{pll}]{Phase-Locked Loop}
\acrodef{pmu}[\textsc{pmu}]{Phasor Measurement Unit}
\acrodef{rtds}[\textsc{rtds}]{Real-Time Digital Simulator}
\acrodef{emt}[\textsc{emt}]{Electromagnetic Transients}
\acrodef{tg}[\textsc{tg}]{Turbine Governor}
\acrodef{avr}[\textsc{avr}]{Automatic Voltage Regulator}
\acrodef{agc}[\textsc{agc}]{Automatic Generation Control}
\acrodef{gps}[\textsc{gps}]{Global Positioning Satellite}

\graphicspath{{./eps/}}
\DeclareGraphicsExtensions{.eps,.png,.pdf}

\def \R {{\rm I\kern -2.2pt R\hskip 1pt}}

\newcommand{\PreserveBackslash}[1]{\let\temp=\\#1\let\\=\temp}

\IEEEoverridecommandlockouts
\begin{document}

\title{Analysis of Wind Energy Curtailment in the Ireland and Northern Ireland Power Systems}

\author{ \IEEEauthorblockN{ Manuel Hurtado\IEEEauthorrefmark{1},
    Taulant K\"{e}r\c{c}i\IEEEauthorrefmark{1},
    \IEEEmembership{IEEE~Member}, Simon Tweed\IEEEauthorrefmark{1},
    Eoin Kennedy\IEEEauthorrefmark{1}, \\Nezar Kamaluddin\IEEEauthorrefmark{1}, and
    Federico~Milano,~\IEEEmembership{IEEE~Fellow}\IEEEauthorrefmark{3}}\vspace*{0.3cm}
  \IEEEauthorblockA{
    \begin{tabular}{cc}
      \begin{tabular}{@{}c@{}}
        \IEEEauthorrefmark{1}
        Transmission System Operator,\\
        EirGrid, plc, Ireland\\
      \end{tabular} &
      \hspace{0.3cm}
      \begin{tabular}{@{}c@{}}
        \IEEEauthorrefmark{3}
        School of Electrical and Electronic Engineering, \\
        University College Dublin, Ireland\\
      \end{tabular} 
    \end{tabular}
  }
  \thanks{M.~Hurtado, T.~K\"{e}r\c{c}i, S.~Tweed, E.~Kennedy are with
    Innovation \& Planning office, EirGrid plc, Ireland; N.~Kamaluddin
    is with Operations office, EirGrid plc, Ireland.  E-mails:
    \{manuel.hurtado, taulant.kerci, simon.tweed, eoin.kennedy,
    nezar.kamaluddin\}@eirgrid.com}%
  \thanks{F.~Milano is with School of Electrical \& Electronic
    Engineering, University College Dublin, Dublin 4, Ireland. E-mail:
    federico.milano@ucd.ie.}%
  \vspace{-2mm}
}

\IEEEoverridecommandlockouts

\maketitle
\IEEEpubidadjcol

\begin{abstract}
  The All-Island power system (AIPS) of Ireland and Northern Ireland
  currently accommodates up to 75\% of system non-synchronous
  penetration (SNSP) (e.g., wind).  These unprecedented levels of
  renewable penetration challenge the operation of the power system.
  The AIPS is not always able to accommodate all of the available
  renewable generation due to binding operational and technical
  constraints.  In this context, this paper analyses wind energy
  curtailment in the AIPS using actual data.  It is found that there
  is a positive correlation between the installed wind capacity and
  curtailment levels, and that the trend is that these levels
  increase.  The paper also shows that the main driver for curtailment
  in AIPS during 2020-2021 was the operational constraint that imposes
  a minimum number of conventional units online (MUON) (80\% of the
  time), with the SNSP limit accounting for less than 20\%.  Other
  system-wide limits, such as rate of change of frequency (RoCoF) and
  inertia are found to have a negligible impact on wind curtailment.
\end{abstract}

\begin{IEEEkeywords}
  Wind Curtailment, dispatch down, low-inertia, power systems.
\end{IEEEkeywords}

\section{Introduction}
\label{sec:intro}

\subsection{Motivation}
\label{Motivation}

Renewable-rich power systems face several challenges such as ensuring
dynamic stability \cite{8450880}, economically maintaining a balance
of supply and demand, and managing local congestion/constraint
\cite{MAI20221981}.  For example, EirGrid and SONI, the transmission
system operators (TSOs) of Ireland (IE) and Northern Ireland (NI),
respectively, are, at times, forced to dispatch down renewables when
there is more power available from renewable sources than can be
accommodated by the grid.  This may be necessary for various reasons,
among others, local congestion (known as ``Constraint'') and
stability driven limits on non-synchronous generation (known as
``Curtailment'').  The term \textit{dispatch down}, thus, is the
result of the combined effect of constraint and curtailment.
This paper focuses on wind energy curtailment in the All-Island
power system (AIPS), which is, to the best of the authors' knowledge,
the largest synchronous power system in the world able to accommodate
up to 75\% of system non-synchronous penetration (SNSP).


\vspace{-2mm}
\subsection{Literature Review}
\label{sec:literature}


Research in renewable and, in particular, wind energy curtailment in
power systems with very high renewable penetration levels is vast
\cite{MILLSTEIN20211749, ZHANG202131,8832216, AGBONAYE2022487}.  One
of the main well-established conclusions in the literature is that
curtailment will be the ``new normal'' for everyday operations
\cite{FREW20211143}.  This is because curtailment is seen both as a
source of flexibility (maintain the supply-demand balance) and
operating reserves \cite{OLSON201449}.  For instance, references
\cite{rebello2020ancillary, 9789420, 7866938} propose to use the
curtailed wind or solar power output to provide ancillary services
such as automatic generation control, ramping and spinning reserve to
the TSO.
In the same vein, references \cite{8085955} and
\cite{abada2018renewable} discuss the use of emerging technologies
such as battery energy storage and data centers to effectively balance
the excess renewable energy output.




With respect to current and expected curtailed volumes of renewable
output, reference \cite{OSHAUGHNESSY20201068} presents an analysis of
solar power curtailment for four key countries namely Chile, China,
Germany, and the United States.  The paper finds that about 6.5
million MWh of solar power output was curtailed in these countries in
2018 and one of the main reasons is attributable to limited
transmission capacity.  Reference \cite{NYCANDER2020942}, on the other
hand, uses a 2025 model of the Nordic power system to estimate the
amount of future curtailment.  The study considers two wind power
scenarios namely 26 and 33 GW and finds curtailment to be below 0.3\%
and 1.7\%, respectively.  The study suggests that the most effective
measures to decrease curtailment is increased transmission capacity,
as well as flexibility of nuclear generation.

Regarding research on wind curtailment in AIPS, reference
\cite{7018991} discusses the effectiveness of compressed air energy
storage (CAES) for mitigating wind curtailment on a 2020 scenario and
shows that a 270 MW CAES plant in conjunction with a 75\% SNSP limit
can reduce wind curtailment levels to 2.6\%.  Using again a 2020
scenario of AIPS, the authors in \cite{MCGARRIGLE2013544} show that
when the SNSP limit is increased from 60\% to 75\% there is a
reduction in wind curtailment from 14\% to 7\%, while for higher SNSP
values wind curtailment is influenced primarily by the inclusion of
transmission constraints.  These conclusions only in part align with
the analysis presented in this paper.

\vspace{-2mm}
\subsection{Contributions}
\label{Contributions}

Though studies on wind power curtailment are common in the literature,
there is a lack of an analysis based on actual data (i.e., the
majority use futuristic scenarios) and most importantly on a
real-world large-scale renewable-based power system such as AIPS.  In
this context, this paper brings the following specific contributions:
\begin{itemize}
\item An analysis of wind energy curtailment based on a real-world
  renewable-based system namely the AIPS.
\item Show that the main driver currently for curtailment is keeping a
  minimum number of conventional units online (MUON) to ensure system
  security and stability.
\end{itemize}

This paper aims at serving as a reference for other TSOs facing very
high levels of renewable energies in the future and motivate
researchers to come up with solutions on how to best manage
curtailment.

\subsection{Paper Organization}

The remainder of the paper is organized as follows.  Section
\ref{sec:policy} discusses the AIPS operational constraints impacting
wind curtailment.  Section \ref{sec:case} provides the methodology
employed in this work to study wind curtailment levels.  Section
\ref{sec:study} presents the results of the curtailment analysis.
Finally, Section \ref{sec:conclu} presents the main conclusions.

\section{Operational Policy Constraints in AIPS}
\label{sec:policy}

EirGrid and SONI have developed various system-wide operational
constraints to ensure system security and stability is maintained at
all times \cite{policy}.  In particular, in order to ensure system
stability, the AIPS has in place four operational constraints namely:
(i) a SNSP limit; (ii) a MUON limit; (iii) a rate of change of
frequency (RoCoF) limit; and (iv) a minimum inertia floor.  These
constraints are explicitly modeled in the scheduling and dispatch
process of the TSOs.  The evolution and relaxation of these
constraints is crucial to achieve the IE and NI targets of 80\% of
annual electricity demand coming from renewables by 2030
\cite{climate, climate1}.  Note that renewables in AIPS are mostly
non-synchronous (e.g., wind) unlike, for example, countries with large
hydro resources (e.g., Norway) or geothermal (e.g., Iceland).  A brief
description of the above constraints is provided below.

\subsection{System Non-Synchronous Penetration}
\label{sec:SNSP}

With increasing level of installed non-synchronous generation capacity
(e.g., wind and solar), and further high voltage direct current (HVDC)
interconnection to the AIPS, it is necessary to measure and limit the
SNSP to ensure secure and prudent operation of the system
\cite{6805233}, \cite{snsp}.  Its formulation is shown in the equation
below:
\begin{align}
  \label{droop} \rm SNSP (\%) = \frac{\rm Wind  \; + Solar \; + HVDC \;  Imports}{\rm System \; Demand + HVDC \; Exports} \, .
\end{align}

\begin{figure}[t!]
  \begin{center}
    \resizebox{0.99\linewidth}{!}{\includegraphics{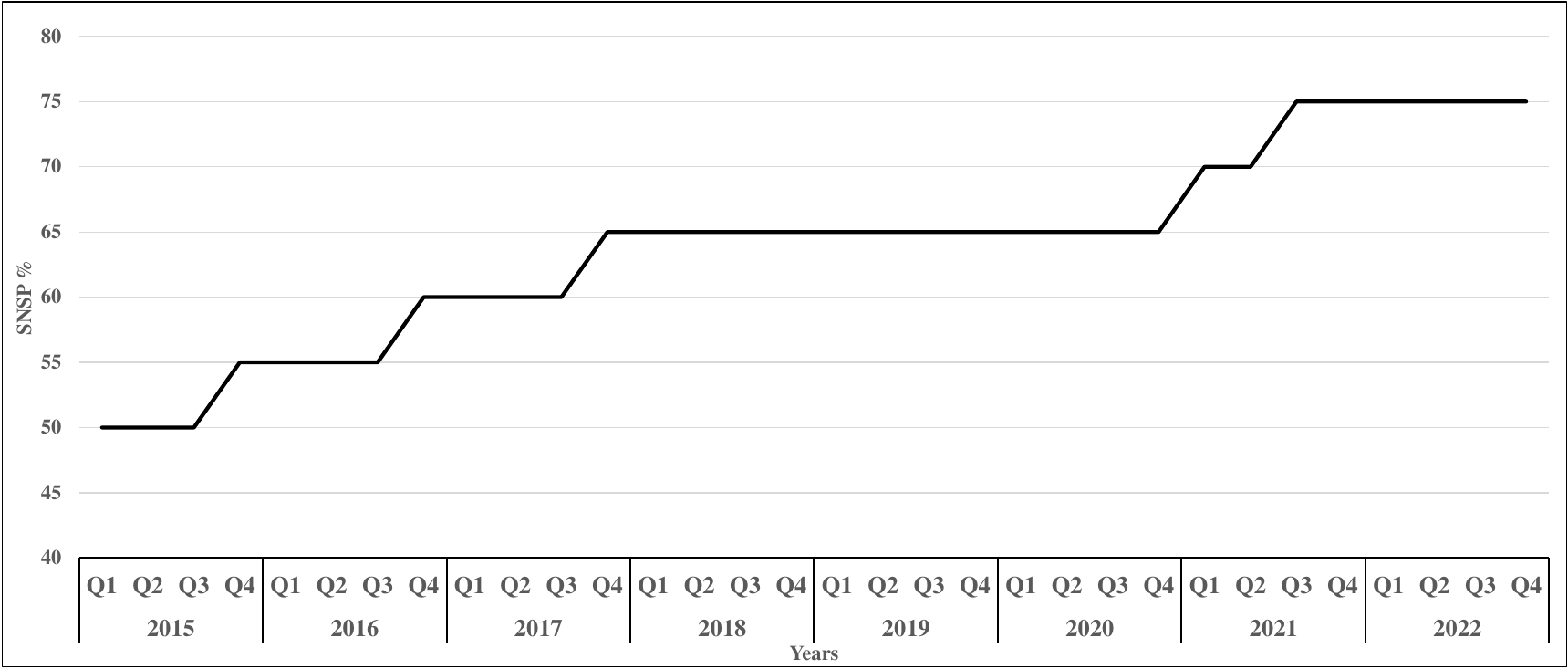}}
    \caption{SNSP evolution over the years.}
    \label{fig:snsp}
  \end{center}
  \vspace*{-0.3cm}
\end{figure}

The SNSP metric was developed around a decade ago based on extensive
studies to provide a single constraint (modeled in the scheduling and
dispatch process of TSOs) that captures the range of issues.  Its
evolution over the years is shown in Fig.~\ref{fig:snsp}.  As it can
be seen, SNSP limit is currently set at 75\%.

\subsection{Minimum Number of Conventional Units Online}
\label{sec:MNON}

The MUON constraint is a system-wide (``catch-all'') constraint.
Currently, it is set to 8 (5 in IE and 3 in NI) large conventional
generation units (from a selected subset of the available generators)
and planning to move to 7 by next year.  It serves to provide
stability for a range of potential system phenomena, such as frequency
stability, voltage, and system strength.  However, there is a need to
relax this constraint to make room for more renewables in the system
and thus achieve both governments' targets.  In this context, EirGrid
and SONI plan to lower this limit to $\leq$ 4 by 2030 \cite{policy}.

\subsection{Rate of Change of Frequency}
\label{sec:rocof}

RoCoF is applied as a constraint in the scheduling and dispatch
process of the TSOs.  Its mathematical formulation is as follows:
\begin{align}
  \label{rocof} \rm RoCoF = \frac{\rm f \times  P_{lost}}{\rm 2 \times (K_{sys} - K_{lost})} \, ,
\end{align}
where f is the system frequency (Hz); $\rm P_{\rm lost}$ is the the
size of the infeed/outfeed lost event (MW); $\rm K_{\rm sys}$ is the
system inertia at the time of the event (MWs); and $\rm K_{\rm loss}$
is the inertia of the unit that caused the imbalance (MWs).  This
constraint is put in place to account for the largest single
infeed/outfeed.  Around a decade ago, the TSOs performed extensive
studies and found out that it would be difficult to ensure stable
system operation with a RoCoF of $\pm$0.5 Hz/s at high SNSP levels
(e.g., $\geq$ 50\%).  As a result, a program of work was followed by
the TSOs, the generator market participants and the distribution
system operators to work towards changing the RoCoF setting on all
assets to $\pm$1 Hz/s.  The TSOs are currently trialling this limit of
$\pm$1 Hz/s.

\subsection{Inertia Floor}
\label{sec:inertia}

In power systems, inertia refers to the ability of the system to
oppose changes in system frequency.  According to \eqref{rocof},
inertia is inversely proportional to RoCoF.  In other words, higher
inertia mean lower RoCoF.  Currently, this limit is set to 23 GWs and
it is planned to lower it to 20 GWs by next year and continue
operating at this limit in the future.  The inertia floor can be met
by a combination of the units that are constrained on by MUON.
Deployment of other low carbon inertia sources (e.g.,
synchronous machines) and potentially inverter-based resources
would contribute to providing inertia in the future.

\section{Methodology of the study}
\label{sec:case}

\begin{figure}[htb]
  \begin{center}
    \resizebox{0.995\linewidth}{!}{\includegraphics{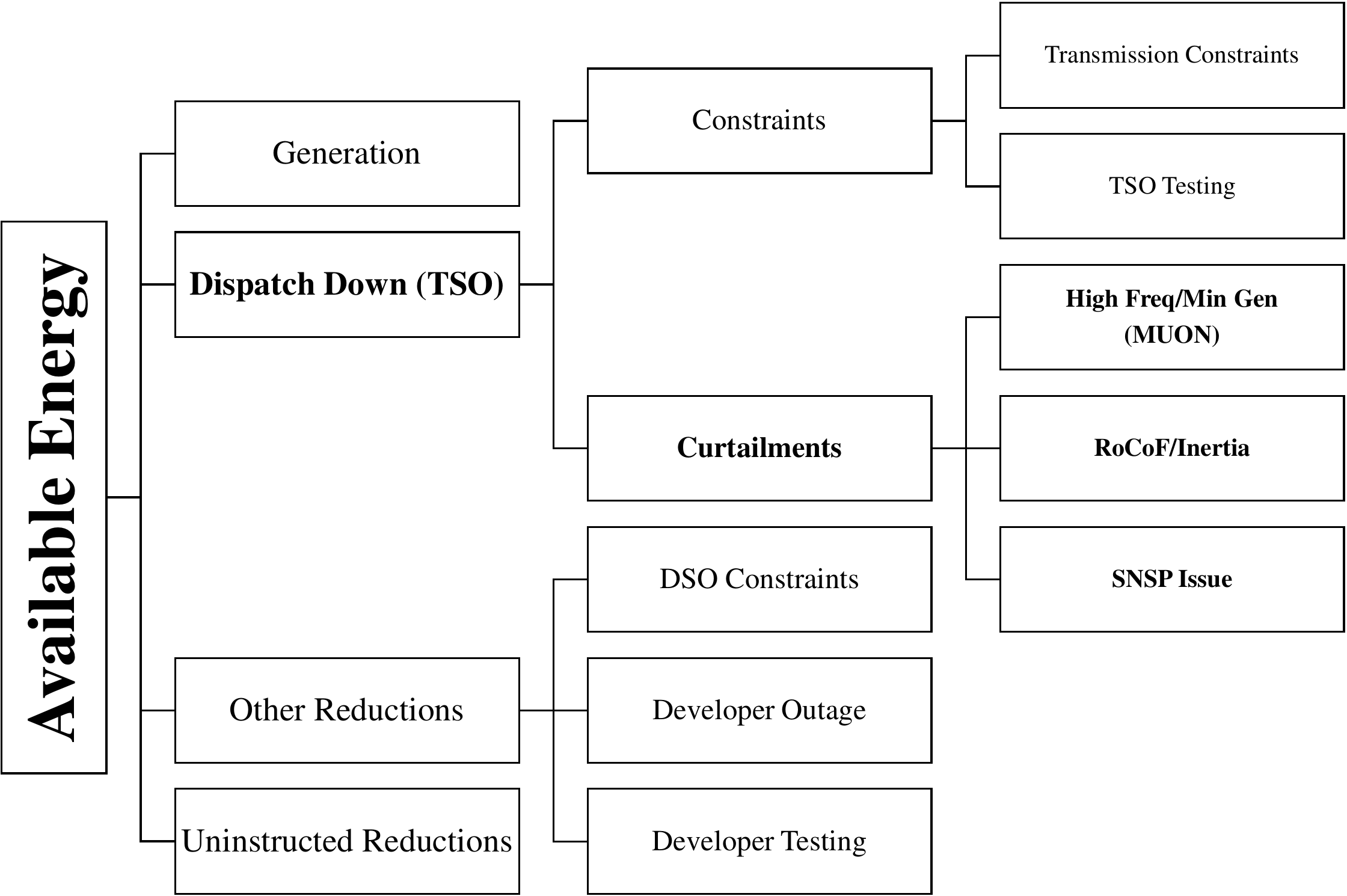}}
    \caption{Overview of dispatch down of wind in the AIPS \cite{dd}.}
    \label{fig:dd}
  \end{center}
  \vspace*{-0.3cm}
\end{figure}

Figure~\ref{fig:dd} shows the categorization used to calculate the
dispatch down of renewables in AIPS, and, in particular, the levels of
wind curtailment.  Dispatch down is composed of two terms namely
constraint and curtailment (i.e., total of constraint and
curtailment).  Note that dispatch down of renewables for reasons of
oversupply (i.e., for energy balancing) are not considered in
Fig.~\ref{fig:dd}.  Dispatch down of wind farms in the AIPS is managed
through the wind dispatch tool implemented in the control centers of
the TSOs.

The control center operators use the tool to issue supervisory control
and data acquisition (SCADA) control set points (i.e., for curtailment
or constraint) to wind farms control systems on a pro-rata basis
(applied regionally for constraint and globally for curtailment)
resulting in the change to their active power output.  These
instructions sent to wind farms are labeled with a reason code.
Specifically, as Fig.~\ref{fig:dd} shows, the reason codes for
curtailment are the following: (i) MUON; (ii) RoCoF/Inertia; and (iii)
SNSP.  It is important to mention here that these reason codes can
occasionally overlap (e.g., SNSP issue and MUON).  However, the
methodology for determining the final reason code is based on the
reason code used for the last instruction.

Further, these instructions are reported in the dispatch down reports
for each wind farm.  Next, we select a typical wind farm dispatch down
report and see when curtailment was active during the year in a 1
minute resolution.  The cases when curtailment is active are then used
to download the respective data from the historical information system
(coming from SCADA and stored in 1 minute resolution).  The interested
reader is referred to \cite{dd} for further information on the
dispatch down methodology in the AIPS.

\section{Case Study}
\label{sec:study}

\begin{figure}[htb]
  \begin{center}
    \resizebox{0.9\linewidth}{!}{\includegraphics{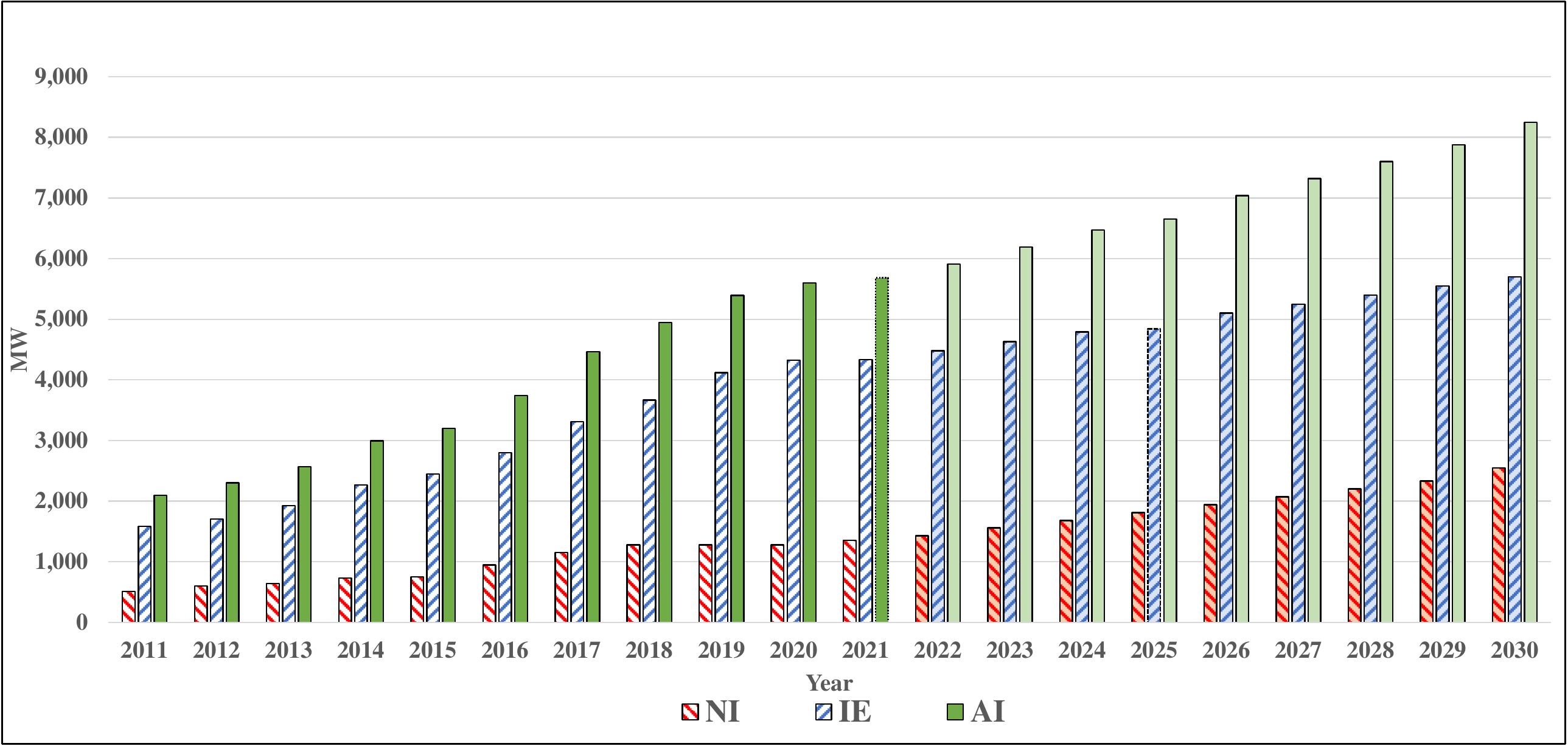}}
    \caption{Evolution of installed and forecasted wind capacity in
      the AIPS \cite{installed}.}
    \label{fig:wind_installed}
  \end{center}
  \vspace*{-0.3cm}
\end{figure}

The AIPS is well-known worldwide for successfully integrating high
levels of non-synchronous renewable generation, particularly wind
power.
Figure~\ref{fig:wind_installed} shows the evolution of installed and
forecasted wind capacity in IE and NI power systems.  Observe the
steady increase in the installed wind in both systems.  The trend is
expected to continue, where significant offshore and on-shore wind
levels are expected by the end of the decade.  At the time of writing,
the AIPS has approximately 5.7 GW installed wind capacity.  This
capacity is slightly below the peak demand of 6.9 GW.
Such high levels of wind capacity present a challenge for TSOs to
ensure system security and stability are maintained (e.g., hard to
control the frequency during low demand and high wind).

\subsection{Historical Levels of Curtailment in the AIPS}
\label{sec:evolution}

\begin{figure}[htb]
  \begin{center}
    \resizebox{0.9\linewidth}{!}{\includegraphics{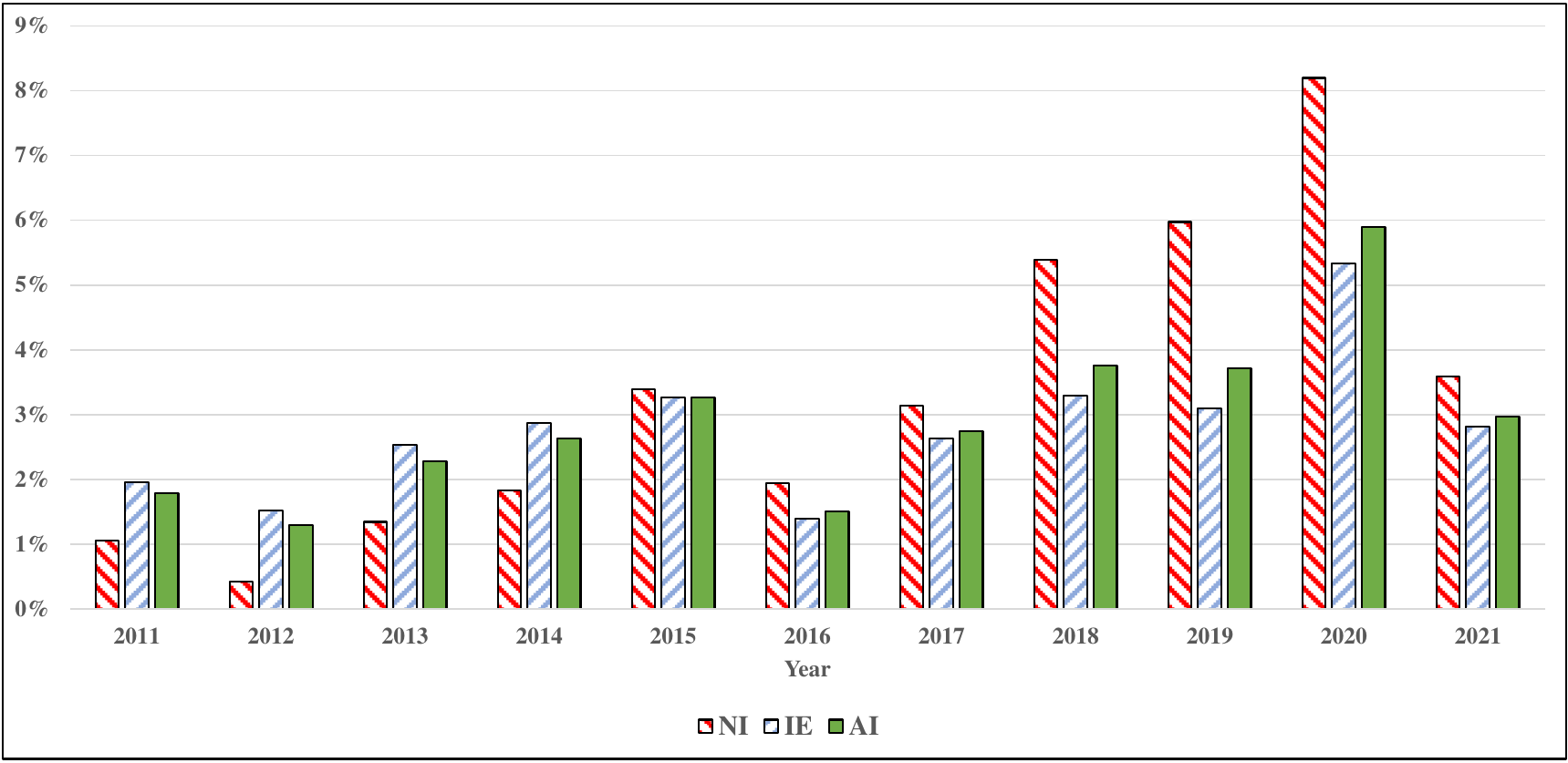}}
    \caption{Historical levels of wind curtailment in the AIPS.}
    \label{fig:curtailment2011}
  \end{center}
  \vspace*{-0.3cm}
\end{figure}

Figure~\ref{fig:curtailment2011} presents the historical levels of
wind curtailment in the AIPS.  Overall, there is a positive
correlation between the installed wind (Fig.~\ref{fig:wind_installed})
and curtailment (Fig.~\ref{fig:curtailment2011}).  An exception is the
year 2021 when wind curtailment has decreased significantly compared
to, for example, the year 2020 (i.e., 2587 number of hours of wind
curtailment in 2020 to 787 hours in 2021).  This is because there was
less wind available in 2021 than in 2020, namely 3,114 GWh and the
SNSP limit was increased to 75\%.

For illustration purposes, Table~\ref{tab:param} shows a breakdown of
wind dispatch down on the AIPS over the last decade, including the
part of constraint.  Overall, curtailment is higher than
constraint.  These results suggest that as the AIPS accommodates
higher and higher levels of wind, significant levels of wind
curtailment are expected if no actions are taken to address them.  As
discussed in section~\ref{sec:policy}, EirGrid and SONI are working
towards relaxing the operational constraints in order to make more
room for wind and, in general, for more renewables in the future.
\begin{table}[t!]
  \centering
  \caption{Wind dispatch-down on the AIPS over the last
decade.}
  \label{tab:param}
  \begin{tabular}{cccccc}
    \hline
    Year  & Constraint & Curtailment & Total Dispatch
Down Levels  \\
    \hline
    2011  & 0.4\% & 1.8\%  &  2.2\% \\ 
    2012 & 0.8\% & 1.3\%   &  2.1\% \\
    2013 & 0.9\% & 2.3\%  &  3.2\%  \\
    2014 & 1.4\% & 2.6\%  &  4.1\% \\
    2015 & 1.8\% & 3.3\%  &  5.1\% \\
    2016 & 1.4\% & 1.5\%  &  2.9\% \\
    2017 & 1.2\% & 2.7\%  &  4.0\% \\
    2018 & 2.2\% & 3.8\%  &  6.0\% \\
    2019 & 4.0\% & 3.7\%  &  7.7\% \\
    2020 & 6.2\% & 5.9\%  &  12.1\% \\
    2021 & 4.4\% & 3.0\%  & 7.4\% \\
    \hline
  \end{tabular}
\end{table}

\begin{figure}[t!]
  \begin{center}
    \resizebox{0.95\linewidth}{!}{\includegraphics{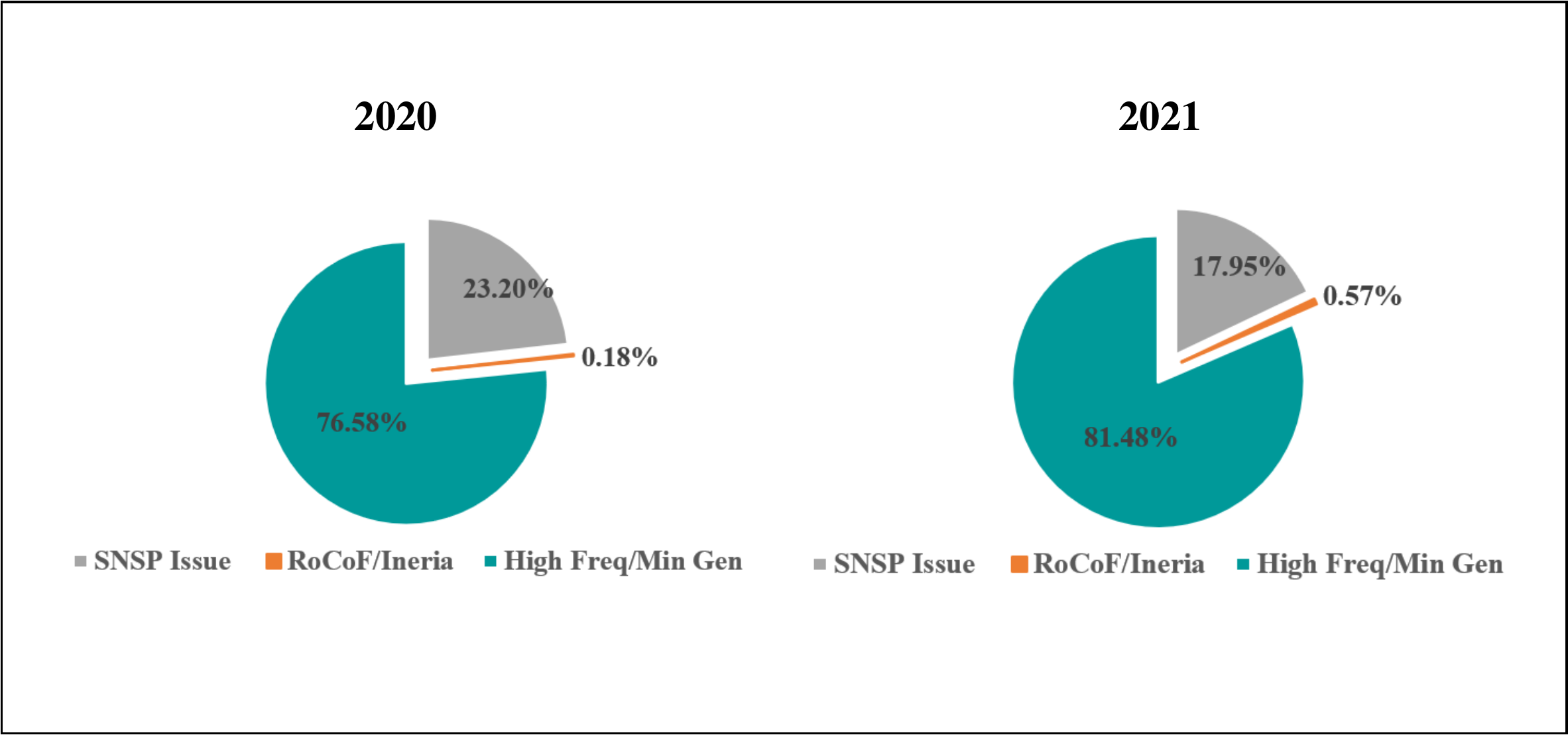}}
    \caption{Reason code (in \%) when curtailment was active.}
    \label{fig:codes}
  \end{center}
  \vspace*{-0.3cm}
\end{figure}

With respect to the reason codes of curtailment, Fig.~\ref{fig:codes}
compares the last two years, namely 2020 and 2021.  One can see that
the reason code percentages are very similar between 2020 and 2021.
The main reason for curtailment during both years is that of MUON with
values of 77\% and 81\%, respectively.  While the SNSP accounts for
23\% and 18\%, respectively.  Note that these figures are different
compared to that predicted by reference \cite{MCGARRIGLE2013544} for
2020.  On the other hand, RoCoF/Inertia appears to account for less
than 1\%.  These results suggest that the AIPS has to relax the MUON
to be able to accommodate more wind in the system.  Indeed, the TSOs
are currently running a procurement process to procure low-carbon
inertia services that will greatly help to reduce the number of
conventional units required to be online \cite{synchcond}.

\begin{figure}[t!]
  \begin{center}
    \resizebox{0.95\linewidth}{!}{\includegraphics{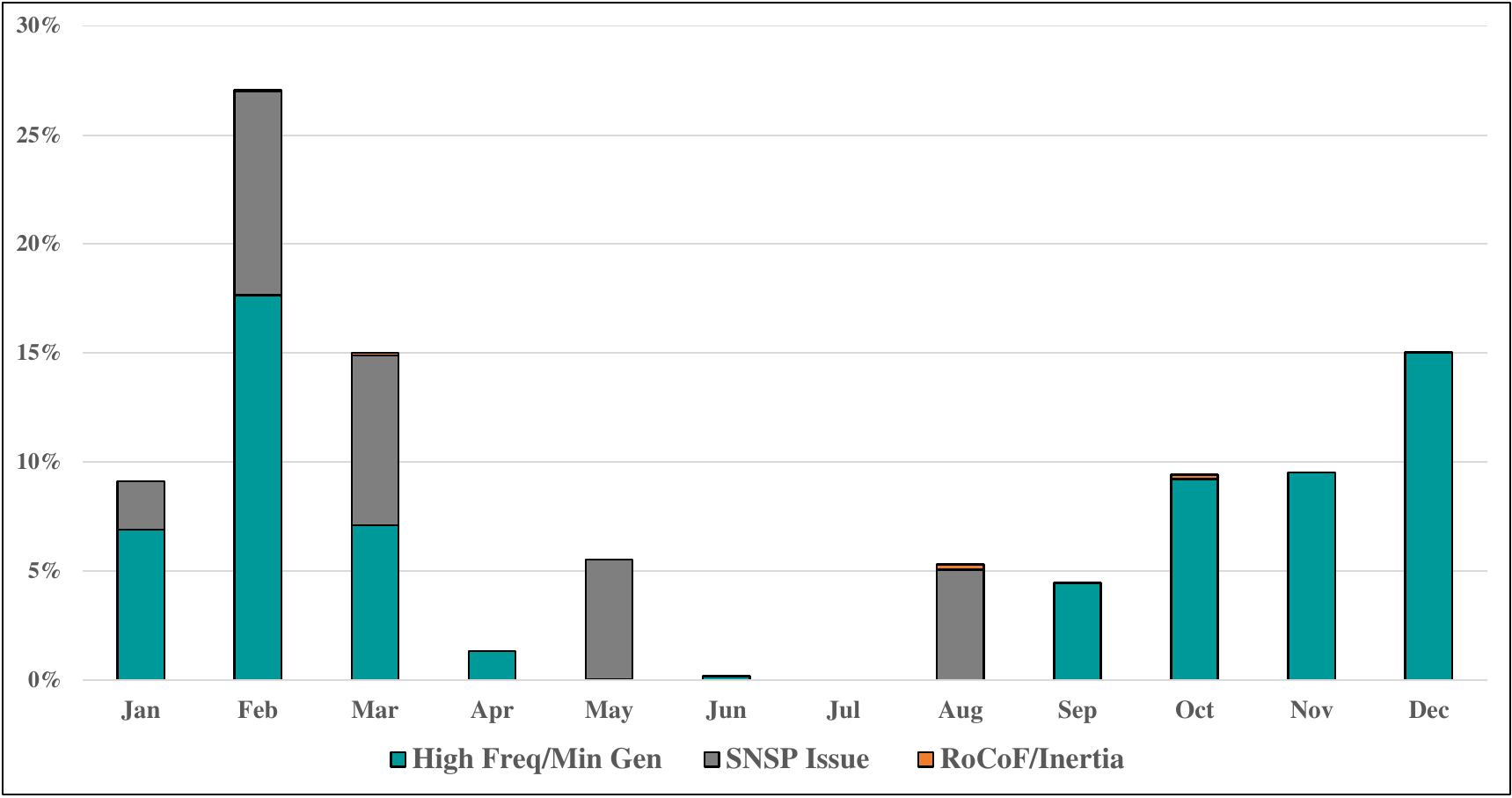}}
    \caption{Curtailment split in months for 2021.}
    \label{fig:months}
  \end{center}
  \vspace*{-0.3cm}
\end{figure}

Figure \ref{fig:months} shows the wind curtailment levels and the
respective reason codes for the  year 2021.  As expected, curtailment
levels are higher during the winter months (i.e., higher wind
availability) and lower during the summer months (i.e., lower wind
availability).  In particular, it is interesting to see that during
February 2021 curtailment was active more than 25\% of the time.  The
figure also shows that again the MUON is, in general, currently the main reason for curtailment.

\subsection{Conventional Generation Ranges While Curtailing Wind}
\label{sec:conventional}

\begin{figure}[t!]
  \begin{center}
    \resizebox{0.95\linewidth}{!}{\includegraphics{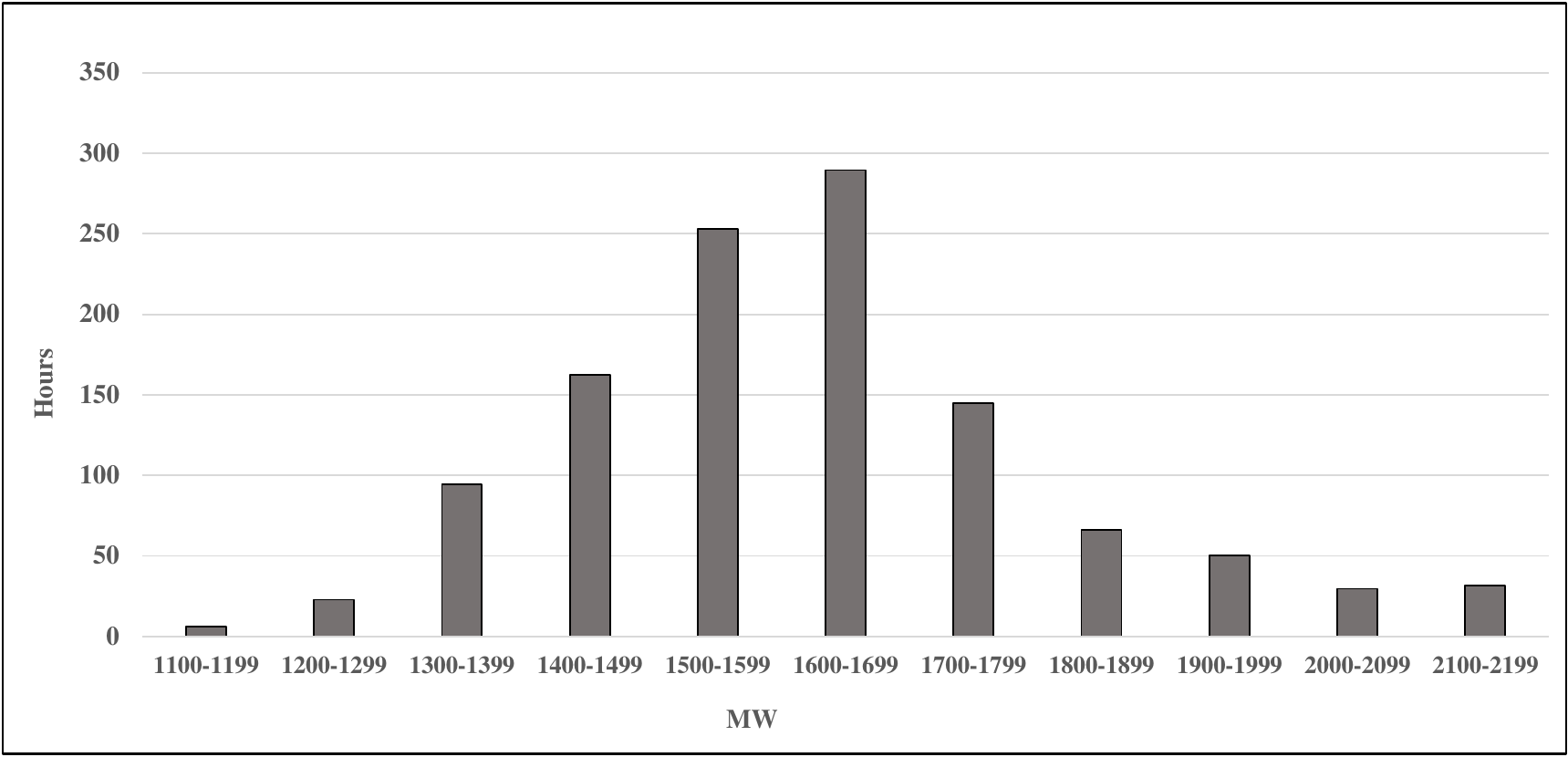}}
    \caption{Conventional generation ranges for 2020 when curtailing wind.}
    \label{fig:conven2020}
  \end{center}
  \vspace*{-0.3cm}
\end{figure}

In this section, we compare the conventional generation ranges (in MW)
while curtailing wind for the last two years, namely 2020 and 2021.
Ideally, before curtailing wind the AIPS should be able to operate
with the MUON (e.g., currently 8) at their minimum levels (i.e., to
make room for renewables).  However, in most cases, this is not
possible because the MUON need to operate above their minimum levels
to provide system services.  Figure~\ref{fig:conven2020} shows the
conventional generation ranges during 2020.  Most of the time, the MW
range of conventional units is between 1400 - 1700 MW.

\begin{figure}[t!]
  \begin{center}
    \resizebox{0.95\linewidth}{!}{\includegraphics{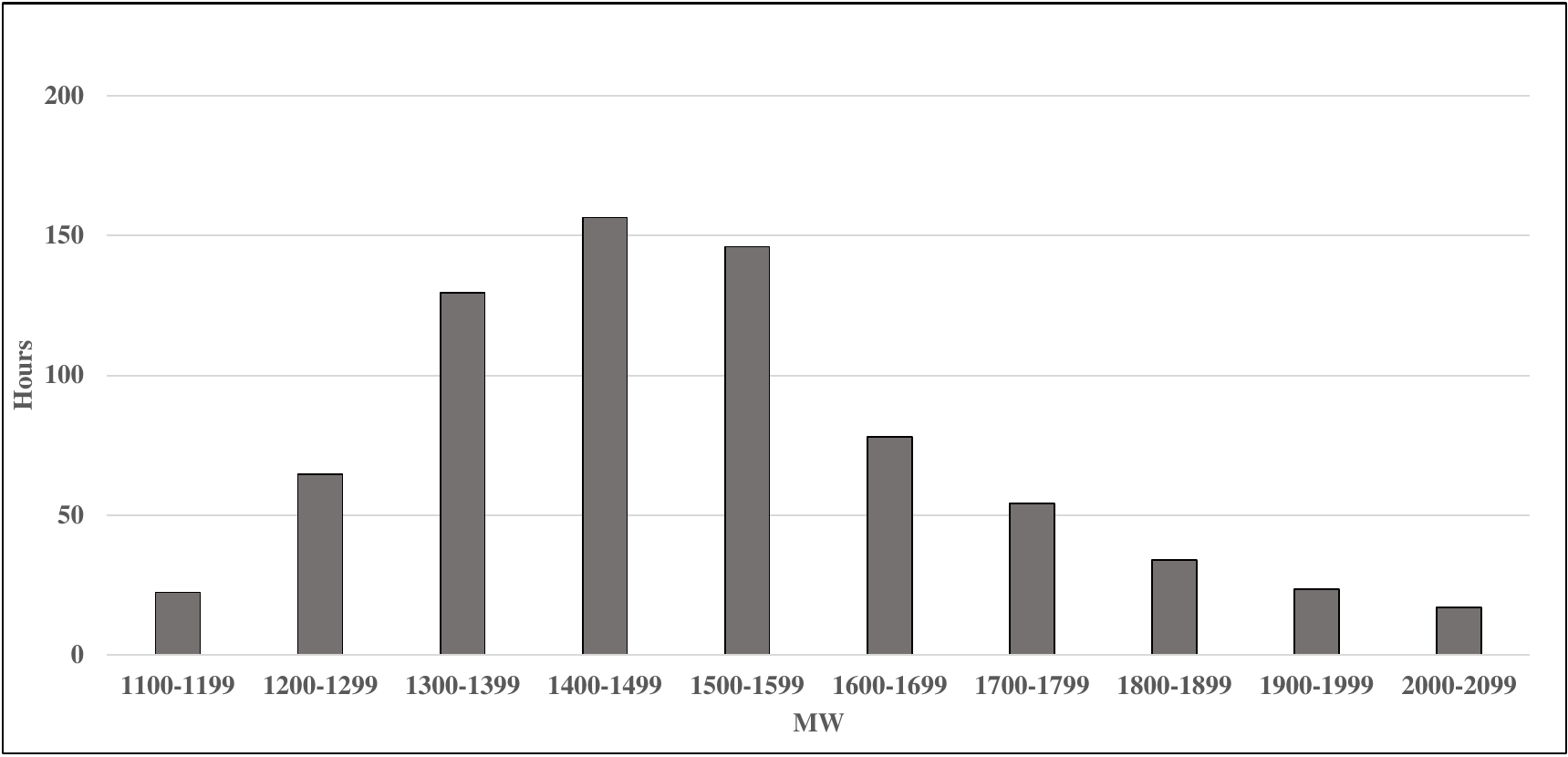}}
    \caption{Conventional generation ranges for 2021 when curtailing wind.}
    \label{fig:conven2021}
  \end{center}
  \vspace*{-0.3cm}
\end{figure}

Figure \ref{fig:conven2021} shows the conventional generation ranges
for 2021, where we can see that this range is between 1300 - 1600 MW
and thus lower compared to 2020.  A contributing factor for this
difference is an operational policy change in 2021, namely the
reduction of negative reserve held on conventional units from 50 MW to
0 MW in IE \cite{policy}.  In the future, reducing the negative
reserve in NI (from 50 MW to 0 MW) will help to reduce the
conventional generation ranges.

\subsection{Interconnector Flows While Curtailing Wind}
\label{sec:interconnectors}

The AIPS is a synchronous island power system with (currently) two
HVDC interconnectors (IC-1 and IC-2) to Great Britain (GB).  The
purpose of this section is to show the interconnector flows while wind
is being curtailed.

\begin{figure}[t!]
  \begin{center}
    \resizebox{0.95\linewidth}{!}{\includegraphics{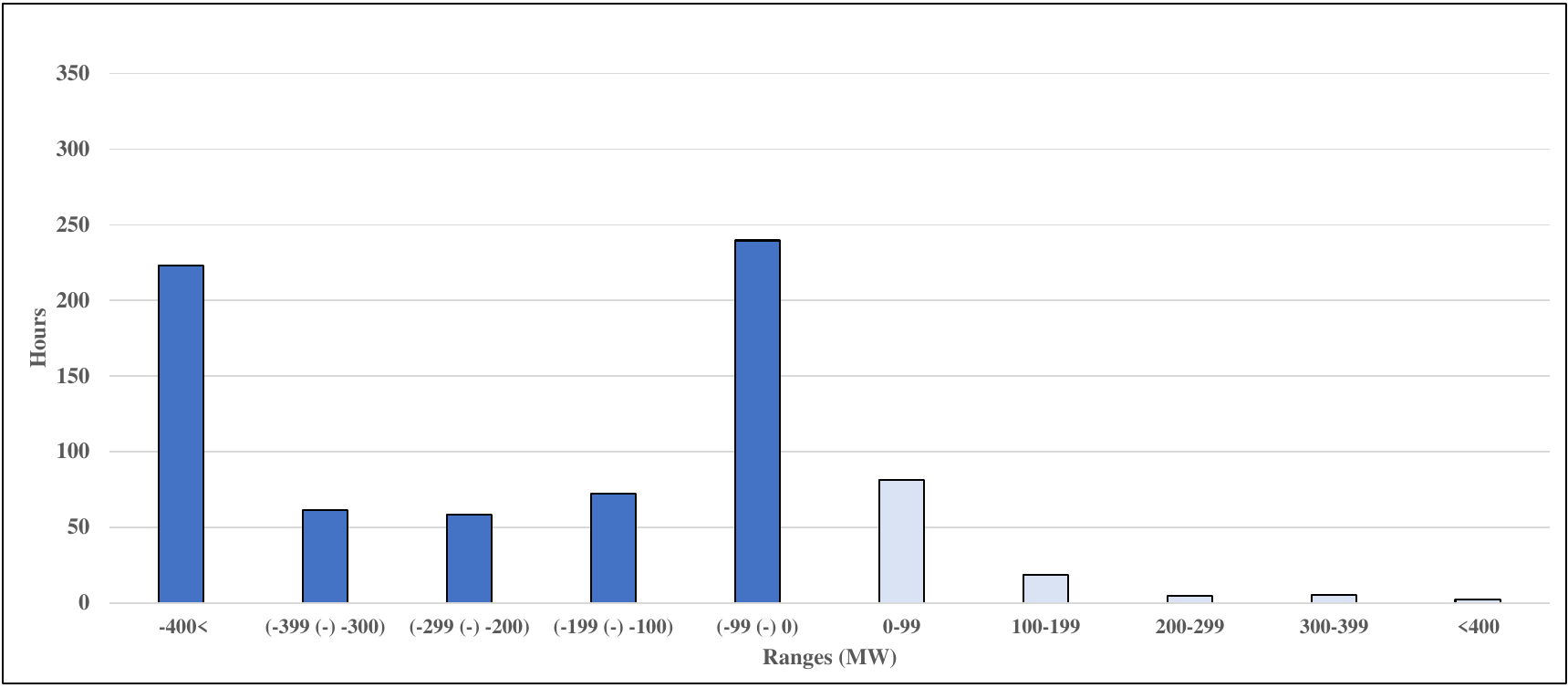}}
    \caption{Interconnector 1 flow during 2021 when curtailment was
      active (negative values indicate export to GB).}
    \label{fig:ewic}
  \end{center}
  \vspace*{-0.3cm}
\end{figure}

\begin{figure}[t!]
  \begin{center}
    \resizebox{0.95\linewidth}{!}{\includegraphics{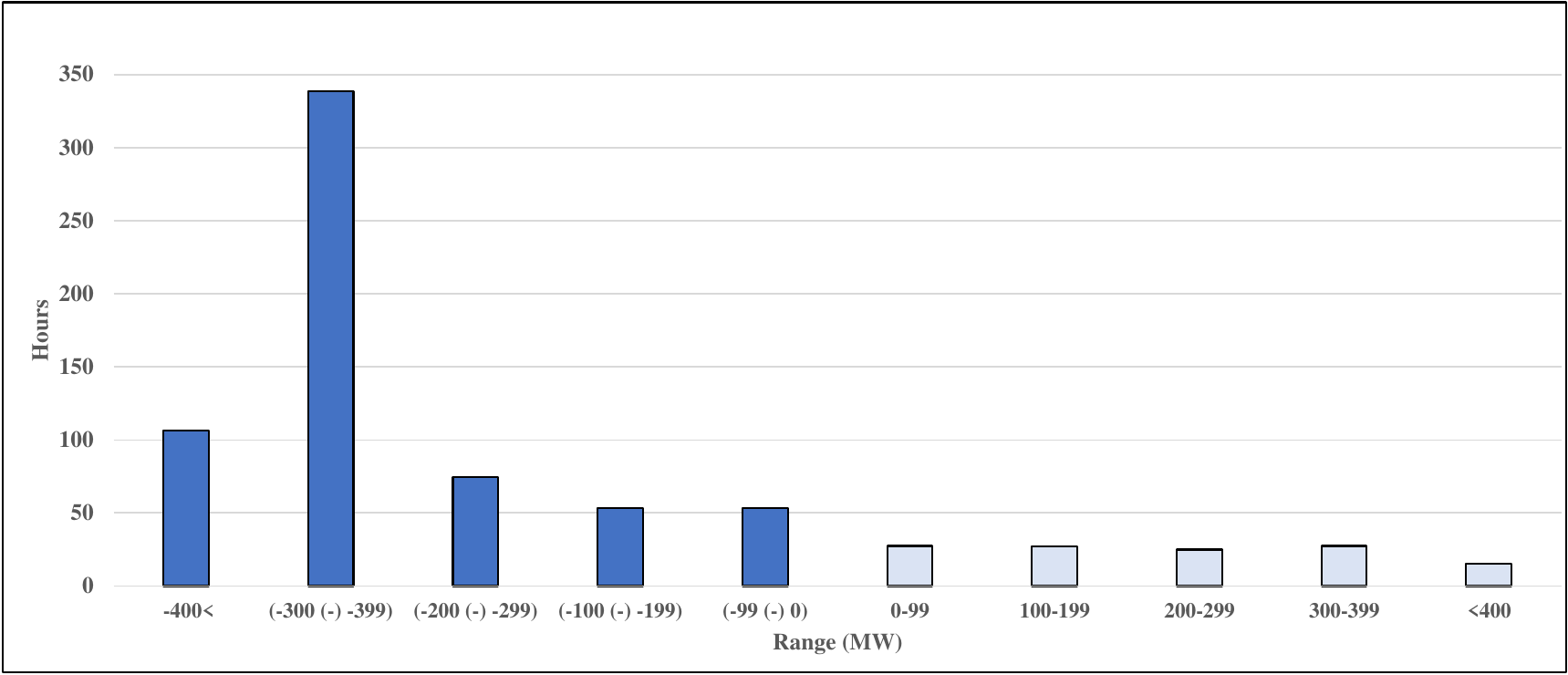}}
    \caption{Interconnector 2 flow during 2021 when curtailment was
      active (negative values indicate export to GB).}
    \label{fig:moyle}
  \end{center}
  \vspace*{-0.3cm}
\end{figure}

Figures \ref{fig:ewic} and \ref{fig:moyle} show the HVDC
interconnector flows for IC-1 and IC-2 during 2021 when wind
curtailment was active.  Figure \ref{fig:ewic} shows that for around
85\% of the time IC-1 was exporting energy to GB.  Further, from that
85\%, the 34\% of that time IC-1 was exporting more than 400 MW.  The
figure also shows that there are other moments where wind is being
curtailed and IC-1 is importing energy from GB (i.e., export headroom
on the interconnector is not used).  These interconnector flows,
however, are determined by the market.  With regard to IC-2, it was
exporting 82\% of the time while curtailment was active in the AIPS.
During that time, 17\% of the time IC-2 is exporting more that 400 MW.
As happened with IC-1, at times IC-2 also appears to import energy
from GB while wind is being curtailed.  Ideally, the latter situations
are not desirable from a wind curtailment point of view.

\section{Conclusions}
\label{sec:conclu}

This paper presents an analysis of wind curtailment in the AIPS.  The
paper shows that there have been significant levels of wind
curtailment over the last decade with the main reason for it during
2020-2021 being the need to keep a MUON to maintain system stability
(80\% of the time).  The SNSP limit, on the other hand, is currently
the bidding constraints for less than 20\% of the time.  While other
system-wide limits such as RoCoF and inertia constraint are found to
have a negligible effect on wind curtailment at this point in time.
These results suggest a need for EirGrid and SONI to continue develop
their operational policy and relax the operational constraints in
order to reduce wind curtailment.  Both TSOs are currently working
towards this direction and plan to publish a 2023-2030 operational
policy roadmap that will detail the pathway to achieve renewable
energy targets and reduce curtailment as much as possible.  The
analysis also indicates that at times there remained significant
interconnector export capacity available when wind was being curtailed
in IE and NI.  However, the market conditions were such that this
capacity was not used.



\end{document}